# Polarization-dependent exciton dynamics in tetracene single crystals

Bo Zhang,[1] Chunfeng Zhang,[1,a)] Yanqing Xu,[1] Rui Wang,[1] Bin He,[1] Yunlong Liu,[1] Shimeng Zhang,[1] Xiaoyong Wang,[1] and Min Xiao[1,2,b)]

[1] *National Laboratory of Solid State Microstructures, School of Physics, and Collaborative Innovation Center of Advanced Microstructures, Nanjing University, Nanjing 210093, China*

[2] *Department of Physics, University of Arkansas, Fayetteville, Arkansas 72701, USA*

**Abstract:**

We conduct polarization-dependent ultrafast spectroscopy to study the dynamics of singlet fission in tetracene single crystals. The spectrotemporal species for singlet and triplet excitons in transient absorption spectra are found to be strongly dependent on probe polarization. By carefully analyzing the polarization dependence, the signals contributed by different transitions related to singlet excitons have been disentangled, which is further applied to construct the correlation between dynamics of singlet and triplet excitons. The anisotropy of exciton dynamics provides an alternative approach to tackle the long-standing challenge in understanding the mechanism of singlet fission in organic semiconductors.

---

a) Electronic mail: cfzhang@nju.edu.cn

b) Electronic mail: mxiao@uark.edu

## I. INTRODUCTION

Singlet fission (SF) in organic semiconductors has recently attracted tremendous attention for its potential to boost the efficiency of solar conversion by creating two triplet excitons from one photo-excited singlet exciton.[1-30] Together with the effect of carrier multiplication in semiconductor nanocrystals,[31-33] SF has been proposed to beat the Shockley-Queisser theoretical limit[34] in single-junction solar cells.[26, 35] In the past few years, remarkable progresses have been made towards this goal in SF-based optoelectronic devices.[3-8] The SF-induced triplet pairs can be harvested by organic electron acceptors[5-7] as well as semiconductor nanocrystals[5, 6, 36, 37] to improve the device performance in solar cells and photodetectors. External quantum efficiency above 100% has already been achieved in a SF-sensitized solar cell device.[4]

In parallel to the impressive development in device engineering, efforts have been made to unravel the physical mechanisms behind such interesting phenomena.[9-27] Most of the experimental and theoretical works have focused on the material family of polyacenes. Tetracene has been regarded as one of these model systems to study the process of SF since 1960s.[9, 10, 12, 14, 17, 18, 22, 29, 30, 38-45] In crystalline tetracene, SF is very efficient although it is a slightly endothermic process. In the latest studies on tetracene with transient optical spectroscopy, the fission rate has been characterized [9, 12, 14, 17, 23] and an intermediate multiexciton state has been identified to mediate the SF process.[17, 18, 21] These results have been inspiring for modeling the nature of SF.[9-11, 17, 21, 39] Nonetheless, some of the key issues are still under debates, such as the effect of coherent coupling,[9, 14, 17] the role of thermal activation for the endothermicity,[12, 17, 23] and the discrepancy on the results recorded from polycrystalline films and single crystals.[23, 40]

Crystalline tetracene has a triclinic structure with herringbone alignment of molecules in ab-plane (Fig. 1(a)).[46-48] Such structure results in anisotropic electronic and optical responses in tetracene crystals, manifested as anisotropic electronic diffusion [49] and polarized optical absorption/emission.[47, 50] Since the transition dipoles are polarized,[47, 50, 51] the excitonic dynamics associated with the excited states should also be anisotropic as pointed out in similar materials.[52, 53] However, this effect has rarely been considered in the available studies on SF dynamics.[38, 40] Here, we report a systematic study on the anisotropic exciton dynamics in tetracene single crystals with polarization-dependent transient absorption (TA) spectroscopy. The spectrotemporal characteristics for singlet and triplet excitons are observed to be strongly dependent on probe polarization. The polarization dependence has then been employed to extract the species of different optical transitions including stimulated emission (SE), ground-state bleaching (GSB) and excited-state absorption (ESA) processes associated with the singlet and triplet excitons, which aids to accurately evaluate the SF dynamics in tetracene crystals. Our method represents an alternative approach to extract the entangled TA components relevant to the dynamics of singlet and triplet excitons, which has been achieved with a complicated algorithm of singular value decomposition in previous literatures.[13, 19, 30] This approach can be naturally extended to exotic photo-physics in a variety of materials governed by oriented transition dipoles.

## II. EXPERIMENTAL

The samples of tetracene single crystals were prepared with a standard method of physical vapor deposition in a quartz tube.[54] The vapor of tetracene was carried by the argon flow (20 sccm) and then deposited onto substrates in the crystal growth zone. Single crystals with thickness of ~1 μm and size up to 5x5 mm$^2$ can be obtained. The crystalline quality was checked by the polarization microscopy and X-ray diffraction. The b-axis of the crystal was approximately determined by the polarization-dependent absorption and emission spectra. We utilized a commercial Ti:Sapphire regenerative amplifier (Libra, Coherent) at 800 nm with a repetition rate of 1 kHz and pulse duration of ~ 90 fs to carry out the TA experiments. A second-harmonic light source at 400 nm was used as the pump beam and an optical parametric amplifier (OperA solo, Coherent) was used to provide a probe beam with tunable wavelength. The pump polarization was set to be parallel to the b-axis while the probe polarization was adjusted by a linear polarizer together with a wave plate. The excitons are initially created at higher energy levels which relax to the lowest singlet excitonic state very rapidly. The pump fluence was set to be ~ 25 μJ/cm$^2$ with the exciton density of ~ $2\times10^{18}$ cm$^{-3}$. We used the balanced detection scheme together with a lock-in amplifier to measure the fractional change in transmission (ΔT/T) with sensitivity better than 10$^{-5}$. The samples were held in a vacuum environment inside a cryostat (MicroCryostatHe, Oxford). All experiments were carried out at room temperature.

## III. RESULTS AND DISCUSSION

### A. Crystalline structure and fluorescence spectra

In tetracene crystals, there are two translationally nonequivalent molecules in each unit cell (Fig. 1(a)). The long axes of the molecules lie close to the c-axis while the short axes of the two adjacent molecules are oriented at an angle of ~ 60°.[51] The dipole moment μ of the lowest singlet transition ($S_0 \to S_1$) in single molecule is oriented along the short molecular axis.[51] Due to intermolecular interactions in crystals, the singlet state ($S_1$) divides into two states originated from symmetric and antisymmetric combination of excitations on the two molecules within the unit, known as Davydov splitting.[55] The resultant lower state is radiative with a dipole polarized along the b-axis, which can contribute to superradiant emission.[50, 51, 56] In the single crystals we studied, the emission spectra show significant polarization dependence (Fig. 1(b)) with an intensity ratio of ~ 15 at room temperature between the polarizations that are parallel and perpendicular to the b-axis, respectively.

**B. Polarization-dependent transient absorption spectra**

We plot the measured polarization-dependent TA spectra in Fig. 2. In general, the TA signal for tertacene consists of multiple components including GSB, SE and ESA.[12, 13, 23, 40] The precise assignments of these components can be complicated in tetracene with entangled dynamics of singlet and triplet excitons. In literature, the transitions of $S_0 \rightarrow S_1$, $S_1 \rightarrow S_n$ and $T_1 \rightarrow T_n$ have been assigned to specific spectrotemporal features.[12, 23, 29, 40] It is well accepted that the fast build-up and decay are features of singlet excitons while the long-lived component is related to triplet excitons. In most previous TA studies, the polarization effects were not considered.[12, 23, 29, 40] In this study, we have observed strong dependences of TA spectra on the probe polarization as shown in Fig. 2(a) and 2(b). The TA spectra obtained with parallel probe polarization consist of three major features including a photo-induced bleaching (PIB) band (2.2-2.3 eV), a broadband photo-induced absorption (PIA) band centered at 1.9 eV, and a long-lived PIA band at 1.5 eV (Fig. 2(a)). When the probe beam is perpendicularly polarized, the PIB band is absent. Besides this difference, the dynamics for the PIA band is also changed.

To understand the polarization dependence, we plot the TA spectra recorded at different delay (Fig. 2(c) and 2(d)) and pump-probe traces at characteristic wavelengths (Fig. 3(a)-3(f)) recorded with parallel and perpendicular polarizations, respectively. The PIB band of 2.2-2.3 eV decays fast, which can be assigned to the SE/GSB associated with the transition of $S_0 \rightarrow S_1$.[12, 23] The polarization dependence is a result of the dipole orientation aligned along the b-axis (Fig. 3(a)).[55] When the probe is perpendicularly polarized, the transition of $S_0 \rightarrow S_1$ becomes very weak. The PIA band is probably induced by ESA associated with the transitions of $S_1 \rightarrow S_n$.[12, 23, 38] The polarization dependence is related to the dipole alignment which depends on the symmetry of $S_1$ and $S_n$ states. Considering the complexity of $S_n$ states, the ESA shows a different polarization dependence from PIB in the spectral range of 2.2-2.3 eV (Fig. 3(d)). The dynamical behaviors of PIB band at 1.9 V are similar for parallel and perpendicular probe beams except for the signal amplitudes.[12]

Next, we consider the PIA band at 1.5 eV. The long-lived feature has been regarded as a characteristic feature of the triplet excitons. The exact assignment has been under debate in previous studies.[12, 23, 40] The onsets of the time-resolved trace consist of a fast instantaneous rise component followed by a slow delayed-rise one.[12, 23] Thorsmolle *et al.* assigned these two components to SF from $S_n$ and $S_1$ states, respectively.[23] The latter process is slow as a result of the endothermicity, i.e., $E_{S_1} < 2E_{T_1}$, where thermal activation might be involved.[23] Recently, Wilson *et al.* reported that the SF process in tetracene is independent of the temperature.[12] The fast component was also ascribed to the ESA associated with the dynamics of singlet excitons.[12] The dynamics of triplet excitons can be reconstructed by eliminating the fast component of singlet dynamics.[12] The polarization-dependent data that we obtained can clearly resolve the above discrepancy. With a parallel probe beam, the onset of PIA process is very fast which starts to decay at an early stage. In contrast, the fast-rise component becomes less important for the onset of the curve recorded with the perpendicular probe beam. Following the slow rise, the PIA remains largely static in the first 500 ps. Such polarization dependence is consistent with Wilson's assignment that the PIA component includes contributions from both singlet and triplet excitons.[12] With different dipole alignments for transitions of $S_1 \rightarrow S_n$ and $T_1 \rightarrow T_n$, the ESA contributions for singlet and triplet excitons become dominant at different polarizations, which can also explain the different results observed on the single crystals and polycrystalline films.[23, 40] In polycrystalline films, Burdett *et al.* found that the long-lived PIA component at 1.53 eV is much weaker than that in single crystals,[40] which can be naturally explained as a result of relatively disordered dipole alignments.

**C. Decomposition of the dynamics for singlet and triplet excitons**

To be quantitative, we recorded pump-probe traces related to the singlet and triplet dynamics with different polarization angles with respect to the b-axis in the tetracene crystal (Fig. 4(a) and 4(b)). We took the early-stage signal (~ 2 ps) probed at 2.21 eV and the long-lived signal (~ 3 ns) probed at 1.59 eV as indicators for singlet and triplet excitons (Fig. 4(c)), respectively. The signal amplitudes are plotted as a function of polarization angle in Fig. 4(c) and 4(d). The degree of polarization can be calculated with the equation of $P = |(I_{//} - I_{\perp})/(I_{//} + I_{\perp})|$, where $I_{//}$ and $I_{\perp}$ represent the signal amplitudes for polarizations to be parallel and perpendicular to the b-axis, respectively. The triplet signal associated with $T_1 \rightarrow T_n$ is found to have a maximum value with polarization perpendicular to the b-axis (Fig. 4(d)) which is about twice as large as that recorded with a parallel polarization. The degree of polarization for $T_1 \rightarrow T_n$ is ~ 0.33 that is comparable to the value studied with steady absorption spectroscopy.[52] It was reported that thermal effect may contribute to the long-lived signal,[12] which seems not to be a major effect here since the polarization dependence is tightly associated with the anisotropic dipole vectors. Nonetheless, this polarization dependence is not the same as that probed at a short wavelength in the work reported by Birech et al. where the maximum signal of triplet absorption was found to be polarized along the b axis.[38] Such a divergence may be caused by the different polarities of the excited levels ($T_n$) involved in the transitions corresponding to the different probe wavelengths used in the experiments.

The signals probed at 2.21 eV vary between positive and negative values when the probe polarization changes, resulting from the superposition of SE/GSB ($S_0 \rightarrow S_1$) and ESA ($S_1 \rightarrow S_n$). Assuming the transition strength to be proportional to $|\mu \cdot \vec{E}|^2$, with μ being the transition dipole and $\vec{E}$ being the vector of electric field,[51] the polarization-angle dependence can then be roughly estimated as $I(\Phi) \propto (\cos^2 \Phi + p)$,[53] where $\Phi$ is the angle between the dipole and electric vector, and p is a parameter related to the degree of polarization (i.e., $P = 1/(1+p)$), respectively. If one assumes the same degree of polarization for SE/GSA and ESA, the signal can be extracted as a sum of the two components from SE/GSB and ESA, i.e.,

$$I(\Phi) \propto I_{SE-GSB}(\cos^2 \Phi + p) - I_{ESA}(\cos^2(\Phi + \Delta\Phi) + p). \qquad (1)$$

By approximately taking the degree of polarization for $S_1$ state to be the same as that of fluorescence, the ESA component is found to be close to 45° ($\Delta\Phi$) with respect to the b-axis (Fig. 4(c)) while the SE component is polarized along the b-axis. The polarization-dependence of singlet signal is in consistence with a recent study on TIPS-pentacene films where the TA signal from single crystalline grains was well characterized.[53] Once the polarization dependences of the singlet and triplet excitons are established, we can assign the spectral profiles of the transitions involved. Roughly speaking, the late-stage signal (~ 3 ns, Fig. 2(d)) probed with the perpendicular polarization can be regarded as the spectral profile of ESA of triplet excitons ($T_1 \rightarrow T_n$); the early-stage signal (~ 2 ps, Fig. 2 (d)) probed with the perpendicular polarization is mainly contributed by the ESA of singlet excitons ($S_1 \rightarrow S_n$); the early-stage signal (~ 2 ps, Fig. 2(c) ) probed with the parallel polarization consists of the ESA of singlet excitons ($S_1 \rightarrow S_n$) as well as the GSB and SE ($S_0 \rightarrow S_1$). The major features of these assignments are consistent with the results from single-wavelength analysis.[12] Moreover, it is noticeable that the signal at 2.45 eV shows the feature of triplet exciton dynamics as reported in literatures,[14, 22, 40] which is likely to be associated to the feature of $T_1 \rightarrow T_n$ observed in tetracene solution.[14, 40]

We can now get a better understanding of the SF process in tetracene concerning the polarization dependences of different processes. From the observed results, the traces recorded at the band of 2.2-2.3 eV are always entangled by the SE/GSB and ESA components. The polarization effect cannot be excluded in analyzing the SF dynamics in tetracene single crystals.[14] We now can take the polarization-dependent part as the SE/GSB component, i.e., $N_S(t) \propto I_{//}^S(t) - I_{\perp}^S(t)$, where $N_S(t)$ is the density of singlet excitons, $I_{//}^S(t)$ and $I_{\perp}^S(t)$ represent the signals recorded at the band of 2.2-2.3 eV with parallel and perpendicular probes (Fig.5(a)), respectively. This is a good approximation since the signals with parallel and perpendicular probes in ESA component cancel each other.

With the dynamics of singlet excitons, we can reconstruct the dynamics of triplet excitons by eliminating the contributions of singlet excitons from transient curves probed at 1.50 eV following Wilson's approach (Inset, Fig. 5(a)).[12] As shown in Fig. 5(b), the reconstructed dynamical behaviors of triplet excitons with the data obtained with parallel and perpendicular probes are basically similar except for the amplitudes. The polarization ratio remains to be nearly time-independent with a value same as that of ESA derived for triplet excitons (Inset, Fig. 5(b)), which confirms the validity of the procedure as the reconstructed curves represent the dynamics of triplet excitons. This can be argued considering different dipole orientations for transitions of $S_1 \rightarrow S_n$ and $T_1 \rightarrow T_n$. Because the signals for singlet and triplet dynamics have significantly different polarization dependences and temporal evolutions, the polarization dependence of PIA signals will change over time if the ESA signals of singlet and triplet excitons are mixed. With the dynamics of singlet and triplet excitons, we can now quantify the SF dynamics. It is generally accepted that SF in tetracene involves intermediate triplet-pair states prior to dissociation to form free triplet excitons, i.e., $S_0 + S_1 \Leftrightarrow 2T_1 \Leftrightarrow T_1 + T_1$.[2, 20, 21] As introduced in Wilson's work, the SF process can be described with a set of rate equations:[12]

$$\frac{dN_S(t)}{dt} = -k_{SF}N_S(t) - k_{SP}N_S(t) - k_{SS}N_S^2(t) + k_{EF}N_{TP}(t), \tag{2a}$$

$$\frac{dN_{TP}(t)}{dt} = k_{SF}N_S(t) - k_{EF}N_{TP}(t) - k_{SR}N_{TP}(t) - k_D N_{TP}(t), \tag{2b}$$

$$\frac{dN_T(t)}{dt} = 2k_D N_{TP}(t) - k_{NR}N_T(t). \tag{2c}$$

$N_S(t)$, $N_{TP}(t)$, and $N_T(t)$ are the populations of singlet, triplet pair and free triplet

excitons; $k_{SP}$, $k_{SR}$, and $k_{NR}$ are the rate constants for spontaneous relaxations of singlet, triplet pair and free triplet excitons; $k_{SF}$ and $k_{EF}$ are the rate constants for SF and exciton fusion processes; $k_D$ is the dissociation rate for triplet pairs; and $k_{SS}$ is the rate constant of singlet-singlet annihilation. The trapping effect of singlet excitons is neglected in tetracene single crystals with good crystalline quality. The channel of triplet-pair fusion is considered as the dominant inverse process of SF, while the triplet-triplet annihilation from free triplet excitons is neglected since it is less efficient than the fusion process in the temporal scale concerned here.[12] The reconstructed triplet signal is proportional to the total density of paired and unpaired triplet excitons, i.e., $2N_{TP}(t) + N_T(t)$. We adopt the rate constants of slow processes ($k_{SP}$, $k_{SR}$, $k_{NR}$, and $k_D$) from the literatures.[12, 29, 40] By fitting the data with equations (2), we obtain the lifetime constants of SF ($k_{SF}^{-1} \approx 83 ps$), excitation fusion ($k_{EF}^{-1} \approx 156 ps$), and SSA ($k_{SS}^{-1} \approx 7.8 \times 10^{-9} cm^3/s$), respectively. These values are in the same levels as the ones reported in literatures.[9, 12, 22, 40] Among the latest reports, the value of SF rate is below that estimated at higher density in Birech's work[38] but slightly above the values reported under weaker excitations.[12, 22] Two possible reasons may cause such slight divergences. First, the polarization effect has been clearly extracted in this work which was not considered in the earlier studies of tetracene films.[12, 22] Second, the SSA rate is comparable to the extracted singlet fission rate, given the pump fluence used in this experiment, which may cause some uncertainty in estimating the rate constant.[16]

## IV. CONCLUSION

In conclusion, we have systematically investigated the SF dynamics in tetracene single crystals with polarization-dependent TA spectroscopy. The polarization-dependent TA signals have been employed to extract contributions from different transitions related to the singlet and triplet excitons. The SF dynamics, i.e., the correlation between the dynamics of singlet and triplet populations, has been accurately evaluated. Our work suggests that the anisotropy of exciton dynamics can be used as a new tool for understanding the detail mechanisms of SF in organic semiconductors. One can employ this approach to study exotic physics governed by oriented transition dipoles in a variety of molecular crystals. Moreover, polycrystalline samples with small crystalline grains can also be adopted for investigation if microscopy technique is utilized to enable high spatial resolution.[53]

## ACKNOWLEDGMENTS

This work is supported by the National Basic Research Program of China (2013CB932903 and 2012CB921801, MOST), the National Science Foundation of China (91233103, 61108001, 11227406 and 11321063), and the Priority Academic Program Development of Jiangsu Higher Education Institutions (PAPD). We acknowledge Dr Vitaly Podzorov, Dr Yuanzhen Chen, and Dr Jonathan Burdett for providing valuable information for crystal growth and sample characterization, and Dr Xuewei Wu for technical assistance.

**Figure 1**

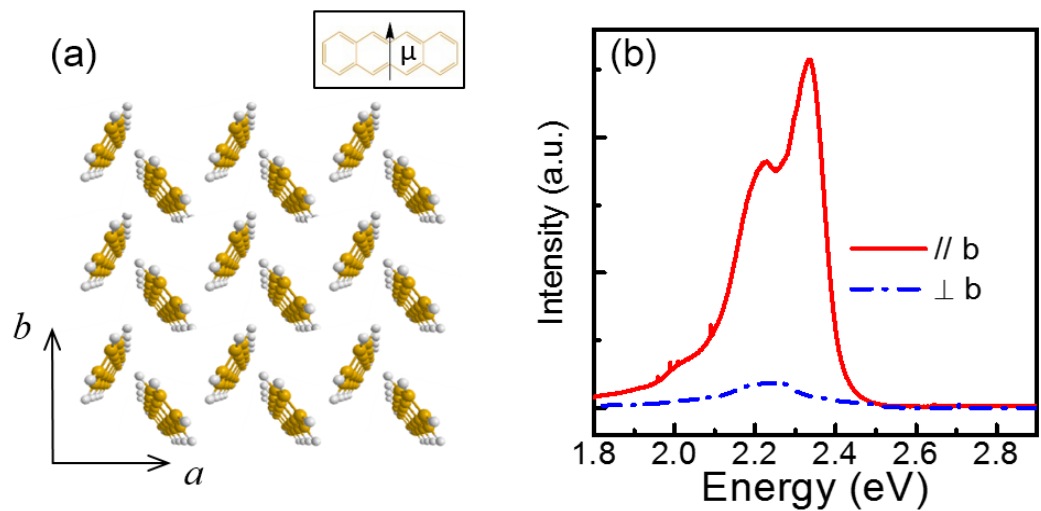

FIG. 1. (a) A c-axis projection of the herringbone structure of tetracene crystal. Inset shows the dipole direction of the lowest singlet transition for single tetracene molecule. The intermolecular interaction results in the dipoles aligned along the b-axis for the radiative transition (Details are available in the text). (b) Fluorescence spectra recorded from the tetracene crystal with parallel and perpendicular polarizations relative to the b-axis, respectively.

**Figure 2**

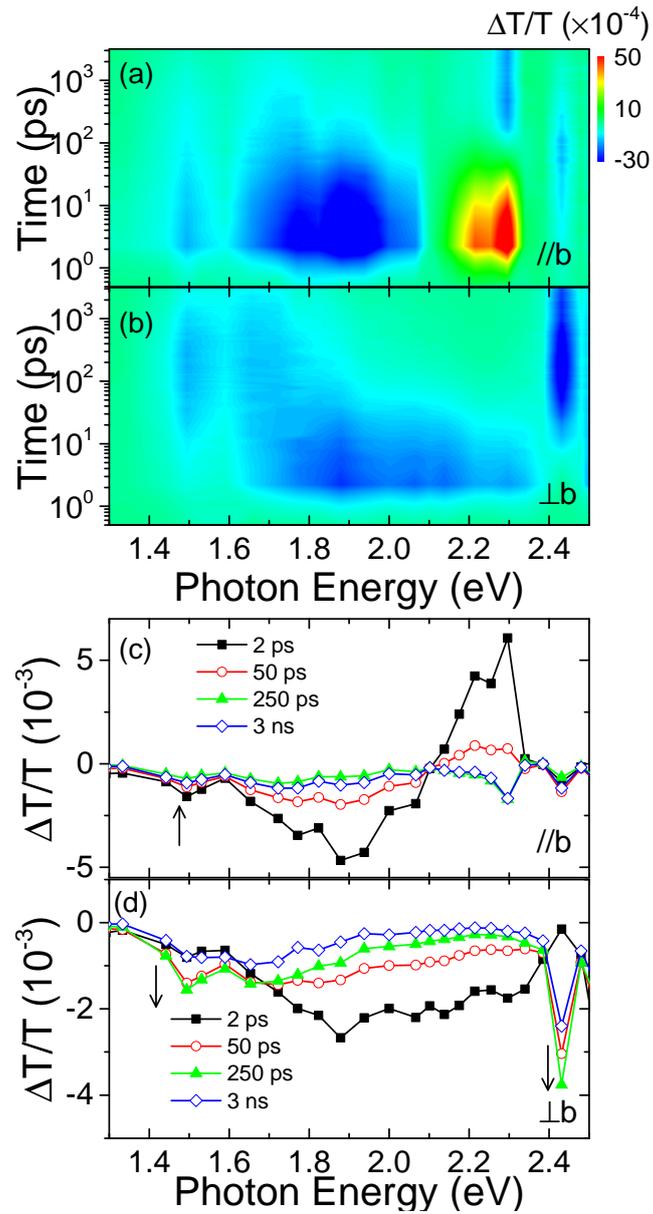

FIG. 2. The TA spectra of tetracene single crystal. The differential transmission variation is plotted as functions of photon energy and delay time with probe polarization parallel (a) and perpendicular (b) to the b-axis, respectively. The TA spectra are shown at different pump-probe delays with probe polarization parallel (c) and perpendicular (d) to the b-axis, respectively. The arrows indicate the temporal evolutions of specific spectral features.

**Figure 3**

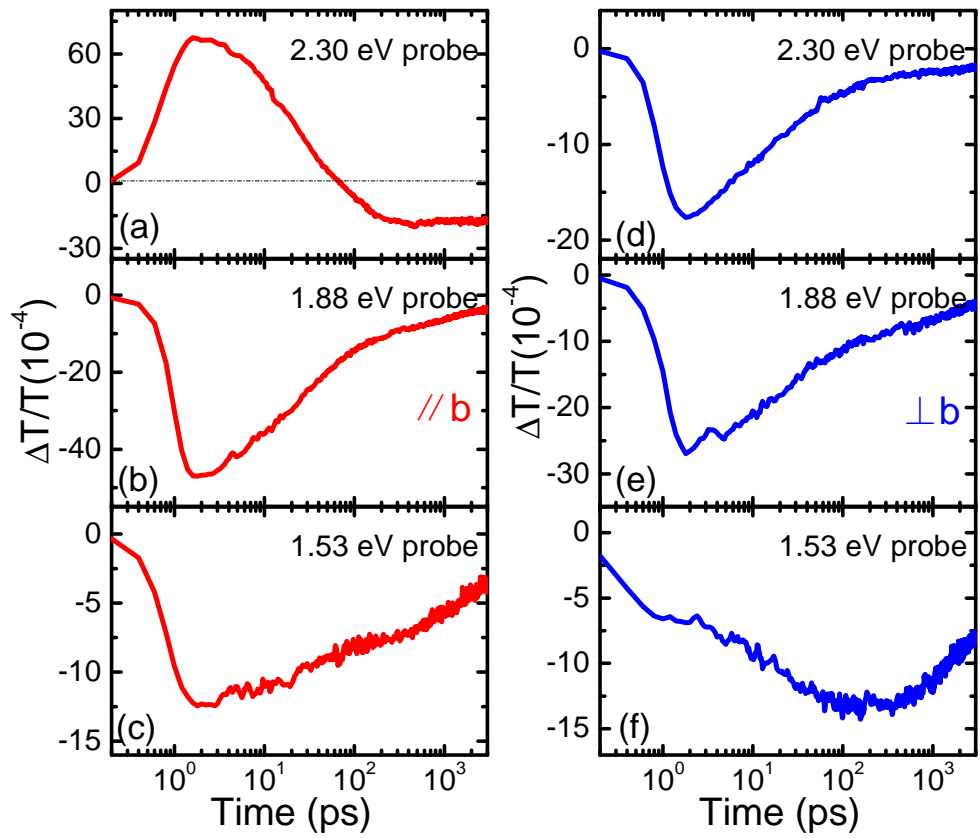

FIG. 3. Pump-probe traces probed with different photon energies with polarization parallel ((a)-(c)) and perpendicular ((d)-(f)) to the b-axis, respectively.

**Figure 4**

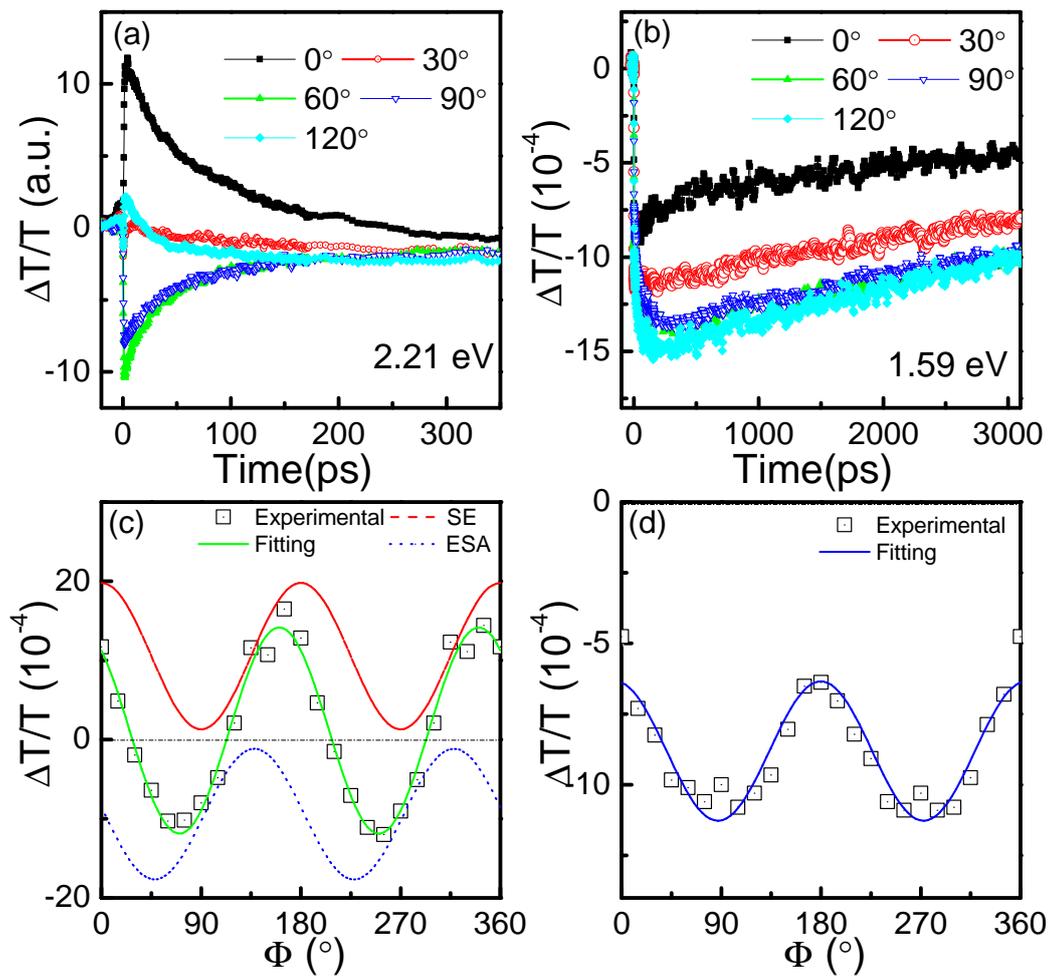

FIG. 4. Polarization-dependent pump-probe traces probed at 2.21 eV (a) and 1.59 eV (b), respectively. The signal amplitudes for singlet (c) and triplet (d) excitons are plotted versus the angle of probe polarization with respect to the b-axis of the crystal. The solid lines are fitting curves with a sine-squared function. The polarization dependence of singlet signal can be extracted in two components of SE and ESA (c).

**Figure 5**

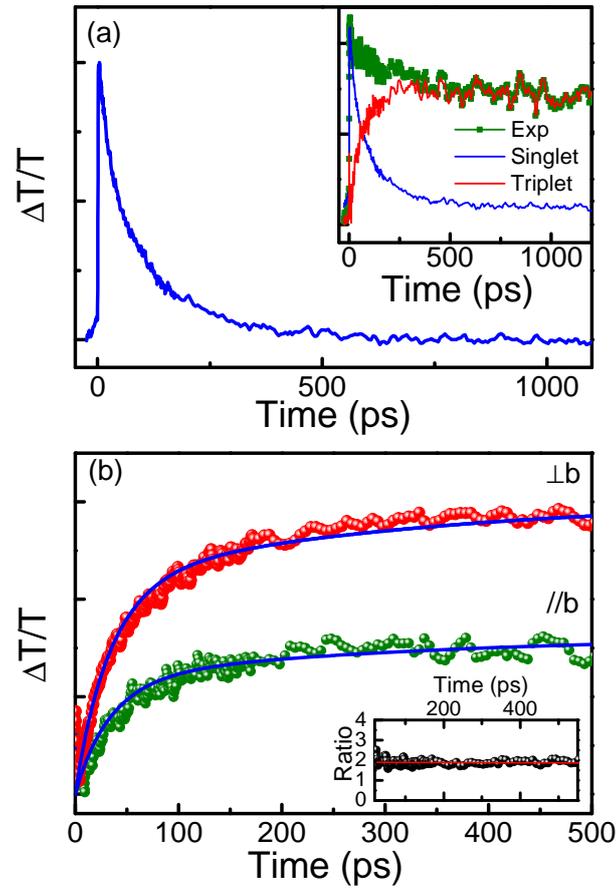

FIG. 5. (a) The difference between signals obtained with parallel and perpendicular probes reveals the dynamics of singlet excitons. Inset shows the reconstruction procedure for triplet dynamics realized by eliminating the singlet contribution from the experimental data recorded at 1.5 eV. (b) The dynamics of triplet excitons reconstructed with the data recorded with parallel and perpendicular probes. The polarization ratio remains static as shown in the inset. The solid lines are fitting curves to the Eq. (1).